\documentclass[aps, pre, reprint, groupedaddress, showpacs, twocolumn]{revtex4-1}

\usepackage{amsmath,amssymb,bm}
\usepackage{graphicx}
\usepackage{color}

\begin{document}

\title{Critical behavior of the two-dimensional icosahedron model}

\author{Hiroshi Ueda$^1$}
\author{Kouichi Okunishi$^2$}
\author{Roman Kr\v{c}m\'ar$^3$}
\author{Andrej Gendiar$^3$}
\author{Seiji Yunoki$^{1,4,5}$}
\author{Tomotoshi Nishino$^{6}$}

\affiliation{$^1$Computational Materials Science Research Team, 
RIKEN Advanced Institute for Computational Science (AICS), Kobe 650-0047, Japan}
\affiliation{$^2$Department of Physics, Niigata University, Niigata 950-2181, Japan}
\affiliation{$^3$Institute of Physics, Slovak Academy of Sciences, SK-845 11, Bratislava, Slovakia}
\affiliation{$^4$Computational Condensed Matter Physics Laboratory, RIKEN, Wako, Saitama 351-0198, Japan}
\affiliation{$^5$Computational Quantum Matter Research Team, 
RIKEN Center for Emergent Matter Science (CEMS), Wako, Saitama 351-0198, Japan}
\affiliation{$^6$Department of Physics, Graduate School of Science, Kobe University, Kobe 657-8501, Japan}

\date{\today}

\begin{abstract}
In the context of a discrete analogue of the classical Heisenberg model, we investigate critical behavior of the icosahedron model, where the interaction energy is defined as the inner product of neighboring vector spins of unit length pointing to vertices of the icosahedron.
Effective correlation length and magnetization of the model are calculated by means of the corner-transfer matrix renormalization group (CTMRG) method. 
Scaling analysis with respect to the cutoff dimension $m$ in CTMRG reveals the second-order phase transition characterized by the exponents $\nu = 1.62\pm0.02$ and $\beta = 0.12\pm0.01$. 
We also extract the central charge from the classical analogue of the entanglement entropy as $c = 1.90\pm0.02$, which cannot be explained by the minimal series of conformal field theory.
\end{abstract}

%\pacs{05.10.Cc, 02.70.-c, 05.30.-d, 71.27.+a}

\maketitle

\section{Introduction}

Statistical models with short range interactions on two-dimensional (2D) regular lattices exhibit no spontaneously symmetry 
breaking at finite temperature, if the symmetry in local degrees of freedom is continuous~\cite{Mermin_Wagner}. 
The classical ferromagnetic XY model 
is a typical example, which has $O(2)$ symmetry, where the thermal average of the magnetization is zero at finite temperature. 
An introduction of discrete nature to local degrees of freedom then induces an order-disorder transition in low temperature,
where the universality class is dependent on the type of discretization. 
The $q$-state clock model, which has $Z_q^{~}$ symmetry, is a well-known discrete analogue of the XY model. 
For the case of $q \le 4$, the clock model exhibits a second-order phase transition described by unitary minimal series of conformal field theory (CFT). 
If $q > 4 $, the clock model has an intermediate critical phase between the high-temperature disordered phase and low-temperature ordered phase~\cite{Elitzur, Nomura, Ortiz, Kumano}, where transitions to the critical phase are of Berezinskii-Kosterlitz-Thouless (BKT) type~\cite{B1, B2, KT}. 
As $q$ increases, the low-temperature ordered phase shrinks, and the $O(2)$ symmetry is finally recovered 
in the limit $q \rightarrow \infty$.

Discretization of the classical Heisenberg model, which has $O(3)$ symmetry, 
is not straightforward, in the sense that there is no established route of 
taking continuous-symmetry limit. A possible manner of discretization is to introduce the polyhedral anisotropies, such as tetrahedral, cubic, 
octahedral, icosahedral, and dodecahedral ones, which correspond to the discrete subgroups of the $O(3)$ symmetry group. 
Let us consider the discrete vector-spin models, where on each lattice site there is a unit 
vector spin that can point to vertices of a polyhedron. The tetrahedron model 
can be mapped to the four-state Potts model~\cite{wu}. For the octahedron model, presence of weak first-order phase transition is 
suggested by Patrascioiu and Seiler~\cite{Patrascioiu}, and afterward is numerically confirmed~\cite{Krcmar}. The cube model can be mapped to 
three decoupled Ising models. 
Patrascioiu {\it et al} reported a second-order transition for the icosahedron 
and dodecahedron models, respectively, which have 12 and 20 local degrees of freedom~\cite{Patrascioiu, Patrascioiu2, Patrascioiu3}. 
For the icosahedron model,  the estimated transition temperatures is $1 / T_{\rm c}^{~} = 1.802\pm0.001$ and its critical indices are $\nu \sim 1.7$ and $\gamma \sim 3.0$, which are inconsistent with the minimal series of CFT. 
By contrast, Surungan {\it et al} gave another estimation $\nu \simeq 1.31$ for the same transition temperature\cite{Surungan}.
However, the system size of Monte Carlo simulations in provious works may be too small to conclude the universality of the icosahedron model.
Finally, a possibility of an intermediate phase is suggested for the dodecahedron model in ~Refs. [\onlinecite{Patrascioiu2}] and [\onlinecite{Patrascioiu3}], whereas a solo second-order transition is suggested in Ref.~[\onlinecite{Surungan}].

In this article, we focus on the critical behavior of the icosahedron model.
We calculate magnetization, effective correlation length and entanglement entropy in the bulk limit by means of the corner-transfer-matrix renormalization group (CTMRG) method~\cite{ctmrg1, ctmrg2}, which is based on Baxter's corner-transfer matrix (CTM) scheme~\cite{Baxter1, Baxter2, Baxter3}. 
An advantage of the CTMRG method is that we can treat sufficiently large system size to obtain the conventional bulk physical quantities. 
Actually, the system size of CTM in this work is up to $10^4 \times 10^4$ sites, which can be viewed as a bulk limit in comparison with (effective) correlation length of the system.
Instead, CTMRG results are strongly dependent on $m$, the number of states kept for the block-spin variables, near the transition point. 
Nevertheless, this $m$-dependence of CTMRG results provides a powerful tool of the scaling analysis with respect to $m$~\cite{fes1, tagliacozzo, pollmann, pivru}, the formulation of which is similar to the conventional finite-size scaling analysis~\cite{Fisher, Barber}. 
The $m$-scaling analysis actually extracts the presence of the second-order phase transition with the critical exponents $\nu = 1.62\pm0.02$ and $\beta = 0.12\pm0.01$.
Another interesting point on the CTMRG  approach is that the classical analogue of the entanglement entropy~\cite{entent} can be straightforwardly calculated through a reduced density matrix constructed from CTMs.
The $m$-dependence analysis of the entanglement entropy also yields the central charge $c = 1.90\pm0.02$, which cannot be explained by the minimal series of CFT.

This article is organized as follows. In the next section, we introduce the icosahedron model, and briefly explain its tensor-network representation and CTMRG method. 
We first show the temperature dependence of the magnetization to capture the nature of the phase transition. 
In Section~III, we apply the finite-$m$ scaling to the effective correlation length, magnetization, and the entanglement entropy. 
Transition temperature, critical exponents, and the central charge are estimated in detail.
The results are summarized in the last section.

\section{Icosahedron model}

\begin{figure}
\includegraphics[width=8.5cm]{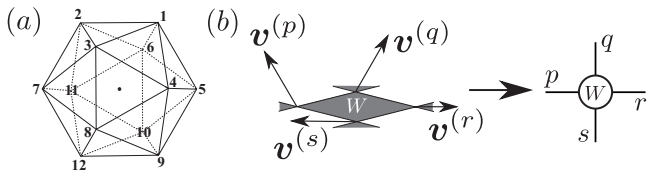}
\caption{
(a) Numbering of the vertices of the icosahedron. 
(b) Local Boltzmann weight in Eq.~(2) defined for a `black' plaquette, 
and its tensor representation. 
}
\label{Fig_1}
\end{figure}

Let us consider the icosahedron model, which is a discrete analog of the classical Heisenberg model.
On each site of the square lattice, there is a vector spin ${\bm v}^{(p)}_{~}\!$ of unit length, which points to one of the vertices of the icosahedron, shown in Fig.~1 (a), where  $p$ is the index of vertices running from 1 to 12.
Figure 1 (b) shows four vector spins ${\bm v}^{(p)}_{~}\!$, ${\bm v}^{(q)}_{~}\!$, ${\bm v}^{(r)}_{~}\!$, 
and ${\bm v}^{(s)}_{~}\!$, around a `black' plaquette, where we have introduced the 
chess-board pattern on the lattice. We have omitted the lattice index of these
spins, since they can be formally distinguished by $p$, $q$, $r$, and $s$, which represent the direction of the spins. 
Neighboring spins have Heisenberg-like interaction, which is represented by the inner product between them. 
Thus, the local energy around the plaquette in Fig.~1 (b) is written as
\begin{eqnarray}
h_{pqrs}^{~} = - J && \left( 
{\bm v}^{(p)}_{~} \! \cdot {\bm v}^{(q)}_{~} + 
{\bm v}^{(q)}_{~} \! \cdot {\bm v}^{(r)}_{~} \right. \nonumber\\
&& +  \left.
{\bm v}^{(r)}_{~} \! \cdot {\bm v}^{(s)}_{~} + 
{\bm v}^{(s)}_{~} \! \cdot {\bm v}^{(p)}_{~}
\right) \, .
\label{Eq_1}
\end{eqnarray}
In the following, we assume that coupling constant is spatially uniform and ferromagnetic $J > 0$. 

\begin{figure}
\includegraphics[width=8cm]{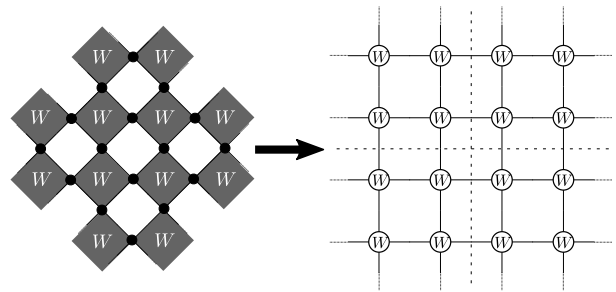}
\caption{Icosahedron model on the diagonal lattice, where $W$ on each `black' plaquette represents local Boltzmann weight
of Eq.~(2). The partition function can be represented by a tensor-network on the square lattice. 
The dashed lines show the division of the system into the quadrants corresponding to CTMs.}
\label{Fig_2}
\end{figure}

We represent the partition function of the system in the form of a vertex model, which can be 
regarded as a two-dimensional tensor network. For each `black' plaquette on the chess-board pattern introduced to the square 
lattice, we assign the local Boltzmann weight
\begin{equation} 
W_{pqrs}^{~} = \exp\biggl[ \frac{h_{pqrs}^{~}}{T} \biggr] \, ,
\label{Eq_2}
\end{equation}
where $T$ denotes the temperature in the unit of Boltzmann constant. 
Note that the vertex weight $W_{pqrs}^{~}$ is invariant under cyclic rotations of the indices. 
Throughout this article we choose $J$ as the unit of energy. As shown in Fig.~1 (b), the weight $W_{pqrs}^{~}$ is 
naturally interpreted as the four-leg tensor, and thus the partition function can be represented as a 
contraction among tensors, as schematically drawn on the right side panel of Fig.~2. 

In Baxter's CTM formulation, the whole lattice is divided into four quadrants~\cite{Baxter1, Baxter2, Baxter3}, as shown in Fig.~2. The partition function of a square-shaped
finite-size lattice is expressed by a trace of the fourth power of CTMs
\begin{equation}
Z = {\rm Tr} \, C^4_{~} \, ,
\label{Eq_3}
\end{equation}
where $C$ denotes the CTM.
Note that each matrix element of $C$ corresponds to the partition function of the quadrant where the spin configurations along the row and  column edges are specified. 
We numerically obtain $Z$ by means of the CTMRG method~\cite{ctmrg1, ctmrg2}, 
where the area of CTM is increased iteratively by repeating the system-size extension and renormalization group (RG) transformation. 
Then, the matrix dimension of $C$ is truncated with cutoff dimension $m$, and under an appropriate normalization, $C$ converges to its bulk limit after a sufficient number of iterations, even if we assume a fixed boundary condition.  
All the numerical data shown in this article are obtained after such convergence. 
The numerical precision of CTMRG results are controlled by the cutoff $m$ for the singular value spectrum $\{\lambda_i\}$ of CTMs with a truncation error $\epsilon(m) = 1-\sum_{i=1}^m \lambda_i^4$. 
The universal distribution of the spectrum  \cite{OHA, cftdistribution} suggests that the asymptotic behavior of $\epsilon(m)$ could be model independent.

\begin{figure}
\includegraphics[width=7.5cm]{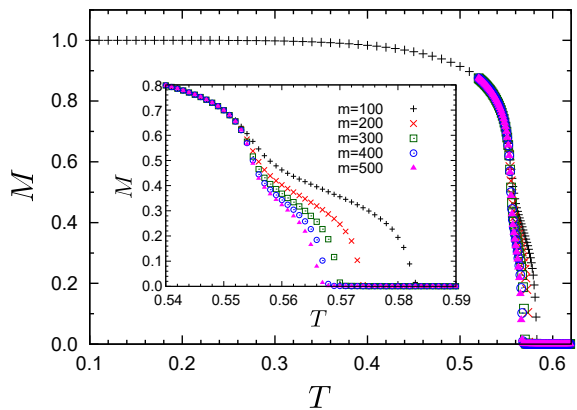}
\caption{(Color online) Temperature dependence of magnetization $M$ for several 
 $m$. The inset: magnified view in the region $0.54 \leq T \leq 0.59$. }
\label{Fig_3}
\end{figure}
In practical computations, we assume the fixed boundary condition, where all the 
spins are pointing to the direction ${\bm v}^{(1)}_{~}\!$ on the boundary of the system.
We define an order parameter as the magnetization $M$ at the center of the system
\begin{equation}
M = \frac{1}{Z} \, \sum^{12}_{s = 1} \, \left( {\bm v}^{(1)}_{~} \! \cdot {\bm v}^{(s)}_{~} \, {\rm Tr}'_{~} 
\bigl[  C^4_{~} \bigr] \right) \, ,
\label{Eq_4}
\end{equation}
where ${\bm v}^{(s)}_{~}\!$ is the vector spin at the center, and 
${\rm Tr}'_{~}\!$ represents partial trace except for ${\bm v}^{(s)}_{~}\!$.
Figure 3 shows the temperature dependence of the magnetization $M$ calculated with
$m = 100$, $200$, $300$, $400$, and $500$. 
The magnetization is well converged with respect to $m$ for $T < 0.55$ or $T> 0.57$, and the result supports emergence of the ordered phase in low-temperature
region as reported by Patrascioiu {\it et al}~\cite{Patrascioiu, Patrascioiu2, Patrascioiu3}.
As shown in the inset, however, the curve of $M$ has the shoulder structure exhibiting the strong $m$ dependence in the region $0.55 <T < 0.57$.

\begin{figure}
\includegraphics[width=7.5cm]{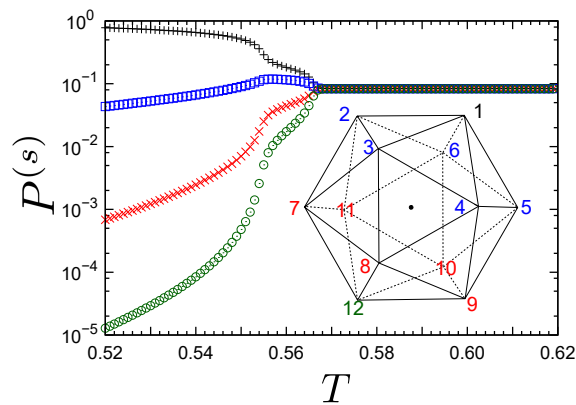}
\caption{(Color online) Probability $P^{(s)}_{~}\!$ of finding ${\bm v}^{(s)}_{~}$ at the center of the system with $m=500$. Plus marks and green circles denote $P^{(1)}_{~}\!$ and $P^{(12)}_{~}\!$, respectively. Blue squares denote $P^{(s)}_{~}\!$ for $s = 2, 3, 4, 5$, and $6$, where these probabilities are the same. 
Red crosses denote $P^{(s)}_{~}\!$ for $s = 7, 8, 9, 10$ 
and $11$. }
\label{Fig_4}
\end{figure}

In order to see the nature of the observed shoulder structure in $M$, we calculate the probability $P^{(s)}_{~}\!$ of finding ${\bm v}^{(s)}_{~}\!$ at the center of the system. Figure 4 shows the temperature dependence of $P^{(s)}_{~}\!$ with $m=500$. 
In the region $T < 0.55$ the probability $P^{(1)}_{~}\!$ is dominant. 
Around $T \sim 0.56$, the value of $P^{(s)}_{~}\!$ for $s = 2, 3, 4, 5$, and $6$ are comparable to $P^{(1)}_{~}\!$, and the sum
$P^{(2)}_{~}\! + P^{(3)}_{~}\! + P^{(4)}_{~}\! + P^{(5)}_{~}\! + P^{(6)}_{~}\!$ is larger than $P^{(1)}_{~}\!$. 
Such a marginal behavior might suggest a possibility of an intermediate (or floating) critical phase, like the two succeeding phase transitions in 
clock-like models~\cite{tobochnik, challa, yamagata, tomita, hwang, brito, baek1, baek2, kumano, 
Kromar2016}. 
We perform the scaling analysis with respect to $m$ to clarify the nature of the phase transition in the next section.

\section{Scaling analysis}

As described above, the calculated results of the magnetization $M$ exhibit the finite-$m$ dependence near the transition point. 
In general, the cutoff dimension $m$ for the CTM introduces an effective correlation length in the critical region~\cite{Tsang, Liu}, which corresponds to a regularization for the infrared divergence.
 By controlling the cutoff $m$, we can systematically analyze the critical behavior in the vicinity of the critical point, which we call the finite-$m$ scaling~\cite{fes1, tagliacozzo, pollmann, pivru}, which shares many aspects in common with the finite-size scaling analysis~\cite{Fisher, Barber}. 

In the scaling analysis, one generally assumes that an observable $A$ with the scaling dimension $x_A^{~}$ 
obeys the following scaling function,
\begin{equation}
A( b, t ) = b^{x_A^{~} / \nu}_{~} \, f_A^{~} \left( b^{1 / \nu}_{~} \, t \right) \, ,
\label{Eq_5}
\end{equation}
where $t = T / T_c - 1$ is the scaled temperature and $b$ is a characteristic length scale intrinsic to the 
system, which basically corresponds to the correlation length. In finite-size scaling analysis, $b$ is replaced 
by the linear dimension of system $\ell$ and then, the scaling function $f_A^{~}$ is extracted by a systematical 
control of $\ell$. Note that the asymptotic forms $f_A( y )^{~} \sim y^{-x_A}_{~}$ for $y \gg 1$ and 
$f_A( y )^{~} \sim {const.}$ for $y \to 0 $ are also assumed in Eq.~(5), in order to reproduce the proper 
scaling law in the bulk limit $\ell \rightarrow \infty$. 

For the finite-$m$ scaling in CTMRG, meanwhile, we introduce a well-controllable length scale through the cutoff dimension $m$, instead of $\ell$. 
After sufficient number of iterations in CTMRG, we have a renormalized row-to-row transfer matrix. 
We can then define an effective correlation length
\begin{equation}
\xi( m, t ) = \Bigl[ \ln\bigl( \zeta_1^{~} / \zeta_2^{~} \bigr) \Bigr]^{-1}_{~} \, ,
\label{Eq_6}
\end{equation}
where $\zeta_1^{~}$ and $\zeta_2^{~}$ are the largest and second-largest eigenvalues of the renormalized row-to-row transfer matrix. 
Note that the unit of length is set as the lattice constant. 
An essential point is that the following scaling relation can be assumed, 
\begin{equation}
\xi( m, t ) \sim m^{\kappa}_{~} \, g\left( m^{\kappa / \nu}_{~} \, t \right) \, , 
\label{Eq_7}
\end{equation}
with the asymptotic forms $g( y ) \sim | y |^{- \nu}_{~}$ for $y \gg 1$ and $g( y ) \sim {const.}$ for $y \to 0$.
Each limit yields the behavior $\xi( m, t ) \sim t^{- \nu}_{~}$ for a finite $t$ under the condition $m^{\kappa}_{~} \gg t^{- \nu}_{~}$, and $\xi( m, t ) \sim m^{\kappa}_{~}$ for a finite $m$ under $m^{\kappa}_{~} \ll t^{- \nu}_{~}$~\cite{fes1, tagliacozzo}. Note that $\kappa$ is an independent scaling dimension, which is characteristic to the matrix-product-state (MPS) description of the eigenvector of the row-to-row transfer matrix. Combining $b \sim \xi( m, t )$ and Eq.~(5), 
we obtain the finite-$m$ scaling formula as
\begin{equation}
A( m, t ) = m^{x_A^{~} \kappa / \nu}_{~} \chi_A^{~}\left( m^{\kappa / \nu}_{~} \, t \right) \, ,
\label{Eq_8}
\end{equation}
where $\chi_A^{~}$ is a new scaling function satisfying $\chi_A^{~}( y ) \sim | y |^{- x_A^{~}}_{~}$ for 
$y \gg 1$. For a finite $t$ under the condition $m^{\kappa / \nu}_{~} \, t \gg 1$, Eq.~(8) reproduces 
$A( m, t ) \sim | t |^{- x_A^{~}}_{~}$, while for a finite $m$ under $m^{\kappa / \nu}_{~} \, t \ll 1$, 
Eq.~(8) gives $A( m, t ) \sim m^{- \kappa x_A^{~} / \nu}_{~}$.

\begin{figure}
\includegraphics[width=7.5cm]{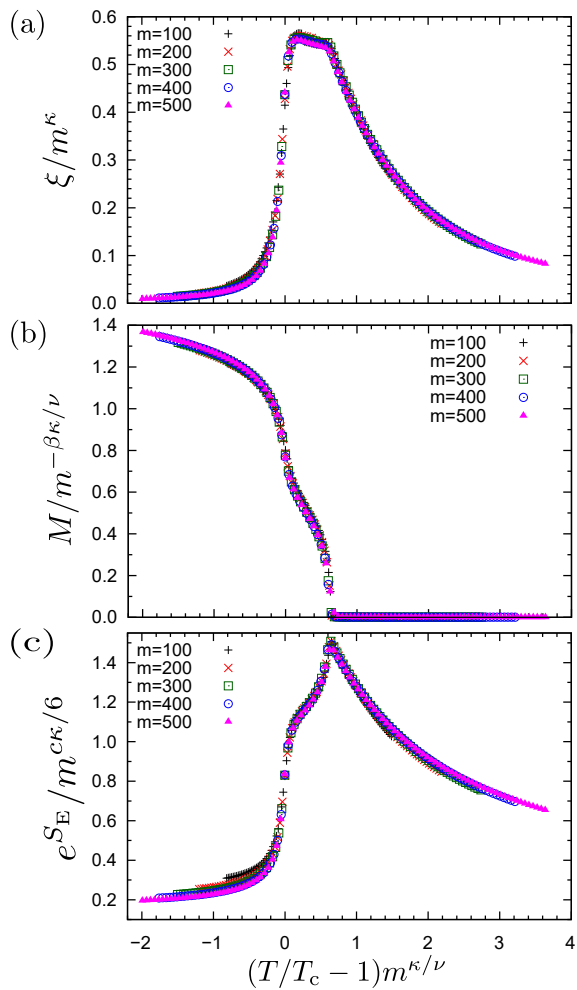}
\caption{(Color online) Finite-$m$ scalings for 
(a) correlation length in Eq.~(6), 
(b) magnetization $M$ in Eq.~(4), and 
(c) entanglement entropy in Eq.~(10).}
\label{Fig_5}
\end{figure}

We apply the scaling analysis to several quantities calculated by CTMRG, with help of the Bayesian analysis for fitting~\cite{Harada}. 
We consider the temperature region 
$0.520 \leq T \leq 0.619$ for $m=100, 200, 300, 400$, and $500$ in the following scaling analysis. 
We first apply the analysis to $\xi( m, t )$ in Eq.~(6) and estimate critical temperature $T_{\rm c}^{~}$, exponents $\kappa$ and $\nu$. Figure 5 (a) shows the scaling plot of $\xi( m, t )$ with the best fit values, $T_{\rm c}^{~} = 0.555048(43)$, $\nu = 1.617(13)$, and $\kappa = 0.8983(17)$, where all data collapse on the scaling function $g$ in Eq.~(7).
The fitting errors in the Bayesian analysis are shown in the round brackets.
If we use the data for $200 \leq m \leq 500$, we obtain 
$T_{\rm c}^{~} = 0.554940(42)$, $\nu = 1.623(13)$, and $\kappa = 0.8830(19)$. Comparing these two fitting results, we adopt $T_{\rm c}^{~} = 0.5550\pm0.0001$, 
$\nu = 1.62\pm0.02$, and $\kappa = 0.89\pm0.02$.
This result of $T_c$ is consistent with the values $T_{\rm c}^{~} \simeq 0.555$ reported by both Patrascioiu {\it et al}~\cite{Patrascioiu, Patrascioiu2, Patrascioiu3} and Surungan {\it et al}\cite{Surungan}.
While, the critical exponent $\nu$ is consistent with the value $\nu = 1.7^{+0.3}_{-0.1}$ in Refs.\cite{Patrascioiu, Patrascioiu2, Patrascioiu3}, but has a discrepancy from $\nu=1.31\pm0.01$ in Ref.~\cite{Surungan}.

On the basis of the above $T_{\rm c}^{~}$, $\nu$, and $\kappa$, moreover, we perform the finite $m$ scaling analysis for the magnetization $M$ shown in Fig.~3.
A particular point is that the shoulder structure in the inset of Fig. 3  directly reflects on the scaling function of Fig. 5(b).
Moreover, such shoulder structures of the scaling functions are consistently observed in Figs. 5(a) and (c).
These behaviors imply that the transition of the icosahedron model is described by a solo second-order transition, unlike to the clock models of $q>4$ where the intermediate critical region emerges.
Using the Basian fitting, then, we obtain $\beta = 0.1293(27)$ for $m=100 \sim 500$ and $\beta = 0.1234(33)$ for $m=200 \sim 500$. 
Taking into account the discrepancy, we adopt $\beta = 0.12\pm0.01$. 
We however think that this value should be improved in further extensive calculations.

In order to obtain additional information for the scaling universality, we calculate the classical analogue of the entanglement entropy. 
The concept of entanglement can be introduced to two-dimensional statistical models through the quantum-classical correspondence~\cite{fradkin, Trotter, Suzuki1, Suzuki2}.  
Then, an essential point is that the fourth power of CTM, which appears in Eqs.~(3) and (4), can be interpreted as a density matrix of the corresponding one-dimensional quantum system~\cite{HU2014}. 
From the normalized density matrix
\begin{equation}
\rho = \frac{C^4_{~}}{Z} \, ,
\label{Eq_9}
\end{equation}
we obtain the classical analogue of the entanglement entropy, in the form of Von Neumann entropy~\cite{vnent1, vnent2}
\begin{equation}
S_{\rm E}^{~} = - {\rm Tr} \, \rho  \ln \, \rho \, .
\label{Eq_10}
\end{equation}

In the context of CTMRG, the following relation
\begin{equation}
S_{\rm E}^{~}( m, t ) \sim \frac{c}{6} \, \ln \, \xi( m, t ) + const.~,
\label{Eq_11}
\end{equation}
is satisfied around the criticality~\cite{Vidal, Calabrese}, where $c$ is the central charge. 
Taking the exponential of both sides of this equation, and substituting Eq.~(7), we obtain
\begin{eqnarray}
e^{S_{\rm E}^{~}}_{~} \sim  a \Bigl[ \xi( m, t ) \Bigr]^{c/6}_{~} 
&=& \, a \Bigl[ m^{\kappa}_{~} \, g\bigl( m^{\kappa / \nu}_{~} \, t \bigr) \Bigr]^{c/6}_{~} \nonumber \\
&=& \, m^{c \kappa / 6}_{~} \, {\tilde g}\bigl( m^{\kappa / \nu}_{~} \, t \bigr) \, ,
\end{eqnarray}
where $a$ is a non-universal constant, and ${\tilde g} \equiv ag^{c/6}$. 
Thus the critical exponent for $e^{S_{\rm E}}_{~}$ is identified as $c \nu / 6$. 

Using $T_{\rm c}^{~}$, $\kappa$ and $\nu$ previously obtained by the finite-$m$ scaling for $\xi( m, t )$, we can estimate the central charge $c$. 
Figure 5 (c) shows the scaling plot of Eq.~(12) for the data of $m = 100, 200, 300, 400$, and $500$.
The central charge is estimated as $c = 1.894(12)$. 
If we exclude the case $m = 100$ for the scaling analysis, we obtain $c = 1.900(15)$. 
Considering the discrepancy between the above values of $c$, we adopt $c = 1.90\pm0.02$.

Here, it should be noted that this value is consistent with the relation
\begin{equation}
\kappa = \frac{6}{ c\bigl( \sqrt{12 / c} \, + 1 \bigr) }\, ,
\label{pollmann}
\end{equation}
which is derived from the MPS description of one-dimensional critical quantum system.~\cite{pollmann}
Substituting $c = 1.90$ and $\kappa = 0.89$ to Eq. (\ref{pollmann}), we actually have $6 / \{ c( \sqrt{12 / c}+1) \} - \kappa = 0.009$, which provides a complemental check of the finite-$m$ scaling in CTMRG.

\section{Summary and discussion}

We have investigated the phase transition and its critical properties of the icosahedron model on a square lattice, where the local vector spin has twelve degrees of freedom. 
We have calculated the magnetization, the effective correlation length, and the classical analogue of the entanglement entropy by means of the CTMRG method. 
The CTMRG results are strongly dependent on $m$, which is the cutoff dimension of CTMs, near the critical point.
We have then performed the finite-$m$ scaling analysis and  found that the all numerical data can be well fitted with the scaling functions including the shoulder structures.
We have thus confirmed that the icosahedron model exhibits the second-order phase transition at $T_{\rm c}=0.5550\pm0.0001$, below which the icosahedral symmetry is broken to a five-fold axial symmetry.
Also, the scaling exponents are estimated as $\nu = 1.62\pm0.02$,  $\kappa = 0.89\pm0.02$, and $\beta=0.12\pm0.01$. 
From the relation between entanglement entropy and the effective correlation length, moreover, we have extracted the central charge as $c = 1.90\pm0.02$, which cannot be described by the minimal series of CFT.
To clarify the mechanism  of such a non-trivial critical behavior in the icosahedron model is an important future issue.

Our original motivation was from the systematical analysis of the continuous-symmetry limit toward the $O( 3 )$ Heisenberg spin.
In this sense, the next target is the dodecahedron model having twenty local degrees of freedom, which requires massive parallelized computations of CTMRG. 
In addition,  it is an interesting problem to introduce the XY-like uniaxial anisotropy to the icosahedron and dodecahedron models;
A crossover of universality between the icosahedron/dodecahedron model and the clock models can be expected, where the shoulder structures of the scaling functions may play an essential role.

\section{Acknowledgment}

This research was  partially supported by Grants-in-Aid for Scientific Research under Grant No. 25800221, 26400387, 17H02931, and 17K14359 from JSPS and by VEGA 2/0130/15 and APVV-16-0186. 
It  was also  supported by MEXT as ``Challenging Research on Post-K computer'' (Challenge of Basic Science: Exploring the Extremes through Multi-Physics Multi-Scale Simulations). 
The numerical computations were performed on the K computer provided by the RIKEN Advanced Institute for Computational Science through the HPCI System Research project (Project ID:hp160262).

\bibliography{reference}

\end{document}